\documentclass[twocolumn,aps,prr,notitlepage,amsmath,amssymb,superscriptaddress,longbibliography,10]{revtex4-1}

\usepackage[utf8]{inputenc}
\usepackage{textcomp}
\usepackage[english]{babel}
\usepackage{graphicx}% Include figure files
\usepackage{dcolumn}% Align table columns on decimal point
\usepackage{bm}% bold math
\usepackage{bm}
\usepackage{amsmath}
\usepackage{verbatim}     %needed for comments
\usepackage[dvipsnames]{xcolor}       %needed for comments
\usepackage[normalem]{ulem}   %needed for strikethrough font
\usepackage{bbold}
\usepackage{physics}
\newcommand{\beq}{\begin{equation}}
\newcommand{\eeq}{\end{equation}}
\newcommand{\bei}{\begin{itemize}}			% Begin itemize environment %
\newcommand{\eei}{\end{itemize}}			% End itemize environment %

\newcommand{\bn}{\mathbf{n}}

\newcommand{\bsigma}{\boldsymbol{\sigma}}

\newcommand{\C}{\mathcal{C}}   %definition of mean chiral displacement
\newcommand{\D}{\mathcal{D}}
\newcommand{\mH}{\mathcal{H}}

\newcommand{\U}{\mathcal{V}_2}
\newcommand{\V}{\mathcal{V}_1}

\usepackage{hyperref}
\hypersetup{%
   pdfpagemode=None, %FullScreen,
   pdfstartpage=1,
   pdfmenubar=true,
   pdftoolbar=true,
   colorlinks = true,
   linkcolor=blue,
   citecolor=blue,
   urlcolor=blue,
   bookmarksopen=false
 } 

\begin{document}

\title{Bulk detection of time-dependent topological transitions in quenched chiral models}
\date{\today}

%%AUTHORS%%
%%
\author{Alessio D'Errico}
\affiliation{Dipartimento di Fisica, Universit\`{a} di Napoli Federico II, Complesso Universitario di Monte Sant'Angelo, Via Cintia, 80126 Napoli, Italy}
\author{Francesco Di Colandrea}
\affiliation{Dipartimento di Fisica, Universit\`{a} di Napoli Federico II, Complesso Universitario di Monte Sant'Angelo, Via Cintia, 80126 Napoli, Italy}
\author{Raouf Barboza}
\affiliation{Dipartimento di Fisica, Universit\`{a} di Napoli Federico II, Complesso Universitario di Monte Sant'Angelo, Via Cintia, 80126 Napoli, Italy}
\author {Alexandre Dauphin}\email{alexandre.dauphin@icfo.eu}
\affiliation{ICFO -- Institut de Ciencies Fotoniques, The Barcelona Institute of Science and Technology, 08860 Castelldefels (Barcelona), Spain}
\author {Maciej Lewenstein}
\affiliation{ICFO -- Institut de Ciencies Fotoniques, The Barcelona Institute of Science and Technology, 08860 Castelldefels (Barcelona), Spain}
\affiliation{ICREA -- Instituci{\'o} Catalana de Recerca i Estudis Avan\c{c}ats, Pg.\ Lluis Companys 23, 08010 Barcelona, Spain}
\author {Pietro Massignan}\email{pietro.massignan@upc.edu}
\affiliation{ICFO -- Institut de Ciencies Fotoniques, The Barcelona Institute of Science and Technology, 08860 Castelldefels (Barcelona), Spain}
\affiliation{Departament de F\'isica, Universitat Polit\`ecnica de Catalunya, Campus Nord B4-B5, 08034 Barcelona, Spain}
\author{Lorenzo Marrucci}
\affiliation{Dipartimento di Fisica, Universit\`{a} di Napoli Federico II, Complesso Universitario di Monte Sant'Angelo, Via Cintia, 80126 Napoli, Italy}
\affiliation{CNR-ISASI, Institute of Applied Science and Intelligent Systems, Via Campi Flegrei 34, 80078 Pozzuoli (NA), Italy}
\author{Filippo Cardano}\email{filippo.cardano2@unina.it}
\affiliation{Dipartimento di Fisica, Universit\`{a} di Napoli Federico II, Complesso Universitario di Monte Sant'Angelo, Via Cintia, 80126 Napoli, Italy}

\begin{abstract}
The topology of one-dimensional chiral systems is captured by the winding number of the Hamiltonian eigenstates. Here we show that this invariant can be read-out by measuring the mean chiral displacement of a single-particle wavefunction that is connected to a fully localized one via a unitary and translation-invariant map. Remarkably, this implies that the mean chiral displacement can detect the winding number even when the underlying Hamiltonian is quenched between different topological phases. We confirm experimentally these results in a quantum walk of structured light.

\end{abstract}

\maketitle
\section{Introduction}
Topological systems are characterized by quantized and global features, known as topological invariants, which are robust to smooth perturbations~\cite{Hasan2010}. Non-zero values of these invariants underlie a variety of remarkable physical phenomena~\cite{Qi2011}, such as the quantization of transport properties in quantum Hall systems~\cite{Klitzing1980,Kane2005,Bernevig1757} or the appearance of robust edge states~\cite{Halperin1982, Hafezi2013, Mancini2015}. 
Promising applications in metrology, spintronics and quantum computation~\cite{VonKlitzing1986,Freedman2002,Nayak2008,Fert2008,Pachos2012} fuelled intense research in materials exhibiting topological order~\cite{Vergniory2019}, leading eventually to the development of artificial topological systems in a variety of physical architectures (e.g.\ cold atom~\cite{Cooper2019}, photonic~\cite{Rechtsman2013, Ozawa2018}, mechanical~\cite{Susstrunk2015}, and polariton~\cite{StJean_2017} systems). 

Topological phases can be classified and related to a specific topological invariant in terms of the system symmetries and dimensionality~\cite{Kitaev2009,Ryu2010}. The simulation and measurement of topological insulators in different quantum simulators is a very active field and different techniques allow one to characterize the topology, relying on the detection of edge states~\cite{Kitagawa2012,  Leder2016, Meier2016, Xiao2017}, center of mass anomalous displacements~\cite{Aidelsburger2015,DErrico2020}, vortex dynamics in reciprocal space~\cite{flaschner2017}, interferometry~\cite{Abanin2013, Atala2013, Flurin2017}, surface scattering~\cite{Barkhofen2017}, time of flight measurements~\cite{Wang2013} and the mean chiral displacement~\cite{Cardano2017, Maffei2018}. 

The investigation of these exotic systems also benefitted from the simulation of topological phases in simple and controllable quantum evolutions known as quantum walks (QWs)~\cite{Aharonov1993}.
Interestingly, quantum walks can be engineered so as to host all topological phases of non-interacting systems in 1D and 2D~\cite{Kitagawa2010}, providing also one of the simplest examples of periodically driven (Floquet) systems~\cite{Kitagawa2010b, Goldman2014, Eckardt2017}. Their study led to a number of theoretical and experimental findings in the context of static and Floquet topological physics, such as the discovery of Floquet anomalous regime~\cite{Kitagawa2012,Cardano2016, Xiao2017,Cardano2017}.

In this article, we focus on one-dimensional (1D) systems with chiral symmetry, whose topological invariant
is the winding number~\cite{Asboth2016,Asboth2013}. 
Some of us showed in Refs.~\cite{Cardano2017, Maffei2018} that such invariant can be accurately detected by measuring the mean chiral displacement (MCD) of a single particle, provided that at a given time this is fully localized on a single lattice site. This method does not require any band filling or the application of an external force, as such it is extremely versatile and has already found application in several experimental scenarios~\cite{Wang2018, Wang2019, Meier2018, Xie2019b,Zhou2019,Bomantara2019,Xie2019}. Here we show that this scheme works also for a much larger class of input states, which may be delocalized over many lattice sites. 
We indeed demonstrate that the input state can be any state that can be mapped to a localized wavefunction via a unitary and translation-invariant operator. Interestingly, such operator can also correspond to the evolution operator associated with an Hamiltonian different from the instantaneous one. As a consequence, we show that in a sudden dynamical transition between two chiral topological phases, hereafter referred to as quench, the mean chiral displacement signals the phase transition and, in the long time limit, eventually tracks the value of the winding number of the post-quench Hamiltonian. After deriving these results, which apply to generic 1D chiral systems, we present an experiment validating our theory in a photonic quantum walk, that implements a Floquet chiral model.

%%%%%%%%%%%%%%%%%%%%%%%%%%%%%%%%%%%%%%%
% Section: theory
%%%%%%%%%%%%%%%%%%%%%%%%%%%%%%%%%%%%%%%

\section{One-dimensional chiral systems}
We consider a system possessing an external discrete degree of freedom, for example its position along a 1D lattice, and an internal degree of freedom, representing the internal state within a given unit cell. These are associated respectively with Hilbert spaces $\mathcal H_e$, spanned by the states $\ket m$ ($m\in \mathbb{Z}$), and $\mathcal H_i$, spanned by $\D$-dimensional vectors, where $\D$ is the internal dimension. 
We assume translational invariance, so that the Hamiltonian can be written as $H=\int \frac{dq}{2\pi}\,  \,\ketbra{q}\mathcal H(q)$, where $q$ is the quasi-momentum defined in the Brillouin zone (BZ) $[-\pi,\pi[$. The Hamiltonian exhibits chiral symmetry if there exists an operator $\Gamma$ which is local, squares to unity and anti-commutes with $H$ (i.e., it acts only on $\mathcal H_i$, and it satisfies $\Gamma^2=1$ and $\Gamma H=-H\Gamma$).  A prototypical example of a chiral Hamiltonian is the Su-Schrieffer-Heeger model~\cite{SSH}, describing the electron effective dynamics in a polyacetylene chain.

One dimensional chiral systems can be characterized by a topological invariant, called the {\it winding number},  that in the reciprocal space reads
\beq
\nu={\rm Tr}\left[\Gamma \mH^{-1} \partial_q \mH\right]
\eeq
and may take arbitrary (positive or negative) integer values. Here, the trace runs over both the 1D external degree of freedom (BZ) and the $\D$-dimensional internal space. 

\section{Mean chiral displacement after a quench} As first demonstrated in Refs.~\cite{Cardano2017, Maffei2018}, the winding of an arbitrary non-interacting 1D chiral model may be conveniently captured by a measurement of the {\it mean chiral displacement},
\beq\label{eq:generalMCD}
\C(t) = \sum_j \langle \psi_j(t)|\Gamma X |\psi_j(t)\rangle,
\eeq
where $X$ is the position operator (acting as $X\ket{m}=m\ket{m}$), and $\{\ket{\psi_j(t)}\}$ $(j=1,\ldots,\D)$ is a set of states such that $\ket{\psi_j(t=0)}=\ket{0}\otimes\ket{\phi_j}$, i.e., which at time $t=0$ are completely localized on the central unit cell of the lattice, and whose internal states $\phi_j$ form a complete basis of the unit cell \footnote{In the case $\D=2$, Eq.\ \eqref{eq:generalMCD} yields twice the value obtained by using the definition provided in Ref.\ \cite{Cardano2017}.}. In the long time limit, one finds $\C(t\rightarrow\infty) \rightarrow \nu$.

Here we generalize these findings by showing that the final result is independent from the early dynamics, provided that this is translation invariant.
In particular, this covers the possibility of studying sudden quenches during the evolution. 
To be specific, let us consider that between time $0$ and time $t_c>0$ the evolution is governed by a translation-invariant Hamiltonian $H_1$. We stress here that $H_1$ does not need to be chiral.
At time $t_c$ a sudden quench is applied to the system, and at later times the evolution is governed by the chiral Hamiltonian $H_2$.
An initially localized state $\ket{\psi_j(t=0)}=\ket{0}\otimes\ket{\phi_j}$ will evolve as
\beq
\ket{\psi_j(t)}=V_2 V_1 \ket{\psi_j(t=0)}= \int \frac{{\rm d}q}{2\pi}\,\U(q)\V(q)\ket{\phi_j},
\eeq
where we have introduced the evolution operators $V_2=\int \frac{dq}{2\pi}\,\ketbra{q}\U(q)=e^{-i (t-t_c) \theta(t-t_c)H_2}$, 
and $V_1=\int \frac{dq}{2\pi}\,\ketbra{q}\V(q)=e^{-i t \theta(t)\theta(t_c-t)H_1}$, with $\theta(t)$ the Heaviside function and $\hbar=1$. Since the position operator $X$ acts as  $i\partial_q$ in Fourier space, the MCD at time $t$  is
\beq
\C=\sum_j\int \frac{{\rm d}q}{2\pi}\, \bra{\phi_j}\V^\dagger \U^\dagger \Gamma\, i \partial_q (\U \V)\ket{\phi_j}.
\eeq
To proceed, we use $\partial_q (\U \V)=(\partial_q \U)\V+\U(\partial_q \V)$ and $V_2^\dagger \Gamma = \Gamma V_2$ (which holds for chiral Hamiltonians). Furthermore, since $\sum_j$ is effectively a trace over the internal space, we use the cyclic property of the trace and the fact that $\V\V^\dagger=\mathbb{1}$, and immediately obtain
\beq\label{eq: MCDterms}
\C= \sum_j\int \frac{{\rm d}q}{2\pi}\,  \bra{\phi_j}
\U^\dagger \Gamma\, (i \partial_q \U) 
+
\V^\dagger \Gamma\, \U^2 (i \partial_q \V)
\ket{\phi_j}.
\eeq

%%%%%%%%%%%%%%%%%%%%%%%%%%%%%%%%%%%%%%%
% Figure: simulations
%%%%%%%%%%%%%%%%%%%%%%%%%%%%%%%%%%%%%%%
\begin{figure}[t]
\includegraphics[width=\columnwidth]{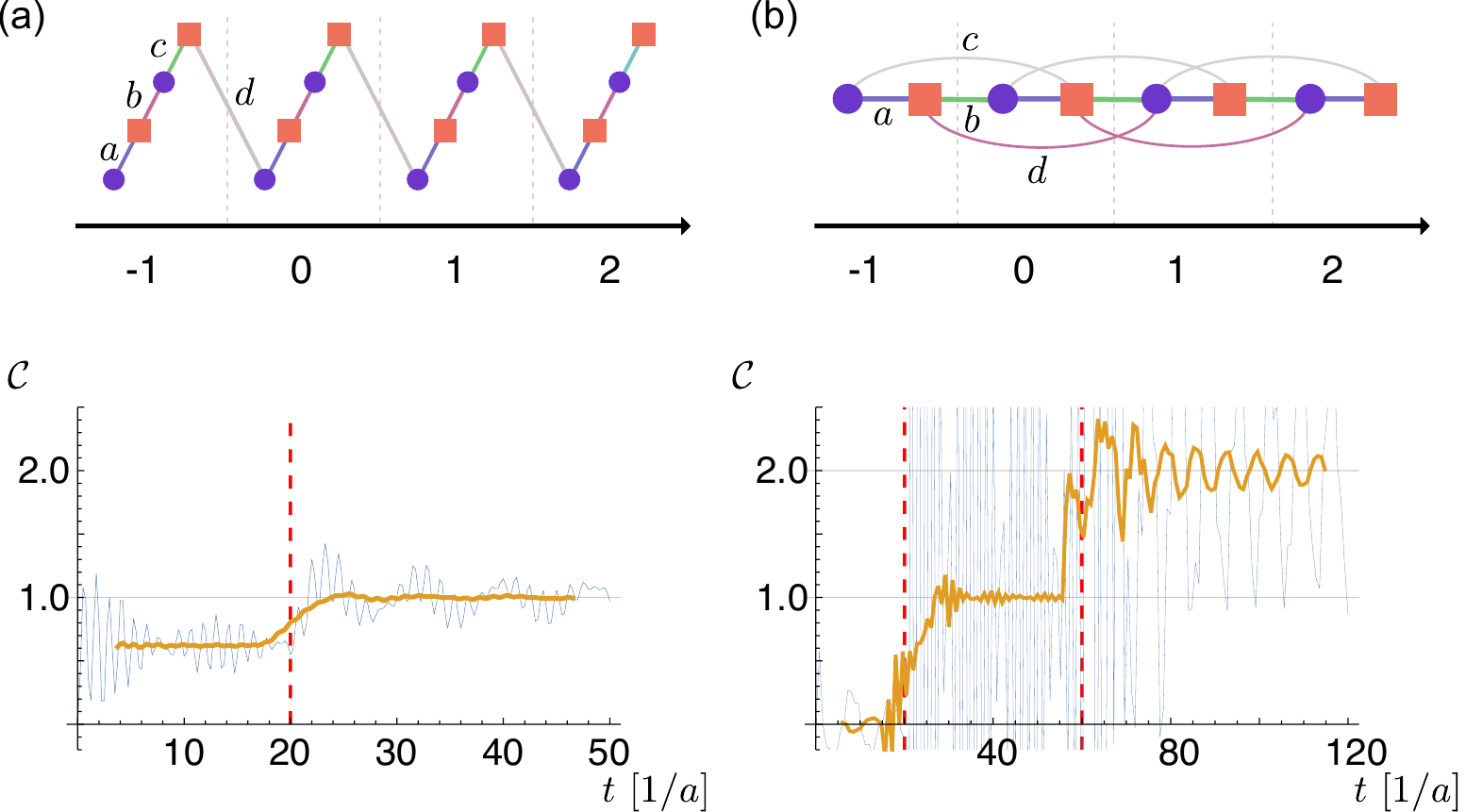}
\caption{{\bf MCD dynamics on simple models.} 
(a) SSH$_4$ model with chiral symmetry broken before $t_c=20/a$, and restored after that [tunnelings: $a=b=c=\frac{d}{2}$, staggering: $\beta=a\,\theta(t_c-t)$].
(b) SSH-LR model quenched from $\nu=0$ to 1 to 2 [tunnelings: $a=b$, $c=0$, and $d/a$ is quenched from 0.5, to -0.5, to 1.5 at times $t_{c,1}=20/a$ and $t_{c,2}=60/a$]. In both panels the thick line is a moving average of the underlying data over a window of duration $10/a$, which removes the fast oscillations.}
\label{fig:simulations}
\end{figure}
%%%%%%%%%%%%%%%%%%%%%%%%%%%%%%%%%%%%%%%

The first operator $\U^\dagger \Gamma\, (i \partial_q \U)$ is identical to the one discussed in Refs.~\cite{Cardano2017,Maffei2018}, and in those works it was shown that it yields a contribution which converges asymptotically to the winding number.
To proceed further, as we are mainly interested in the behavior of the MCD at long times, we will restrict ourselves to $t>t_c$, such that only $\U$ is time-dependent.
We notice that $\U^2=e^{-2i (t-t_c) \theta(t-t_c)\mathcal{H}_2(q)}$, where $\mathcal{H}_2(q)$ is the quasi-momentum space representation of $H_2$. The second operator in the integrand of Eq.~\eqref{eq: MCDterms} therefore generates contributions of the form $\int {\rm d}q\, f(q)e^{i\,t\, g(q)}$, with $f$ and $g$ smooth functions of $q$. These are rapidly oscillating terms with amplitude $\sim1/\sqrt{t}$ and zero mean, as guaranteed by the stationary phase formula~\cite{wong2001}.
In conclusion we obtain
\beq\label{eq:main}
\C
\stackrel{t\gg t_c}{=} \nu 
+O\left(\frac{1}{\sqrt{t}}\right),
\eeq
where $\nu$ is the winding number of the chiral Hamiltonian $H$ generating the dynamics after the quench.
Equation \eqref{eq:main} is the main finding of this paper. Although deceivingly simple, this has a series of remarkable consequences. 
A first one is the convergence of the MCD to the winding number of the Hamiltonian of an initially localized state on any cell of the lattice (as may be seen by taking $\V$ to be an instantaneous translation).
A second one is the possibility of using the MCD as a topological marker even in presence of multiple quenches during the evolution.

Let us now numerically illustrate our finding by two concrete examples of quenches, based on tight-binding models introduced in Ref.\ \cite{Maffei2018}. 
First, we consider the tight-binding model SSH$_4$ sketched in Fig.~\ref{fig:simulations}(a), which has internal dimension $\D=4$. For times $0<t<t_c$ we include in the Hamiltonian $H_1$ a staggering term, which explicitly breaks the chiral symmetry. The MCD signal therefore rapidly oscillates, and settles around a non-quantized value. At time $t_c$ we remove the chiral-breaking term, and soon after the MCD signal is observed to converge to an integer value, which corresponds to the winding number of the model.
In a second example we analyze the SSH-LR sketched in Fig.~\ref{fig:simulations}(b). This is a variant of the SSH model which includes longer-ranged tunnelings, and admits values of the winding number larger than 1. Here we quench the Hamiltonian twice, and after each quench the MCD is shown to converge smoothly to the expected winding number within that phase. 

Closely following the demonstration in Ref.~\cite{Maffei2018} it can be further shown that, when $\D=2$, chiral symmetry ensures that the MCD is equal to twice the expectation value of the operator $\Gamma X$ over a single arbitrary state localized within a single unit cell at $t=0$. 
Similarly for $\D>2$ the trace over the internal space may be replaced by  twice the sum over the $\D/2$ states from a single sublattice, provided the operator $V_1$ does not mix sublattices.

%%%%%%%%%%%%%%%%%%%%%%%%%%%%%%%%%%%%%%%
% Section: experiments
%%%%%%%%%%%%%%%%%%%%%%%%%%%%%%%%%%%%%%%
\section{Quenches in a photonic quantum walk} We verify our theoretical findings in a 1D chiral quantum walk, realized by engineering light propagation through a sequence of suitably patterned birefringent optical elements~\cite{Cardano2015,DErrico2020}. Specifically, as described in greater details in Ref.\ \cite{DErrico2020}, we encode position states $\ket{m}$ (of the particle undergoing the QW) into Gaussian optical modes carrying a quantized amount of transverse wavevector $m\Delta k_\perp$. We keep the lattice spacing $\Delta k_\perp\ll k_z$, where $k_z$ is the longitudinal component of the wavevector, so that these modes remain confined along the $z$ axis. Coin states (that is the internal degree of freedom) instead are mapped into left and right circular polarizations, referred to as $\left\{\ket L,\ket R\right\}$. At each timestep $t$ \footnote{With a slight abuse of notation, we use the same variable $t$ for both the continuous time and discrete timesteps.}, the system evolution is determined by a combination of two transformations: a unitary operator $W$ performing a rotation of the polarization state only, and a translation operator $T$ \cite{Cardano2015}. We implement $W$ by using a quarter-waveplate (QWP) oriented with its fast axis parallel to the optical table. In the basis of circular polarizations, we have
\beq\label{eq:W}
W=
\frac{1}{\sqrt 2}\left( {\begin{array}{cc}
   1 & i \\
   i & 1 \\
  \end{array} } \right).
\eeq
The translation operator $T$ is implemented by liquid-crystal polarization gratings, referred to as $g$-plates \cite{DErrico2020}, whose action is described by the operator
\beq\label{eq:gplates}
T(\delta)\equiv
\left( {\begin{array}{cc}
   \cos(\delta/2) & i \sin(\delta/2) \hat t \\
   i \sin(\delta/2) \hat t^\dagger & \cos(\delta/2) \\
  \end{array} } \right),
\eeq
where $\hat t$/$\hat t^\dagger$ are the (polarization-independent) left/right translation operators, acting as
$\hat t\ket{m}=\ket{m-1}$ and $\hat t^\dagger \ket{m}=\ket{m+1}$, and  $\delta$ is the plate optical retardation, which can be tuned by a voltage applied to the cell~\cite{Piccirillo2010}.

%%%%%%%%%%%%%%%%%%%%%%%%%%%%%%%%%%%%%%%
% Figure: setup
%%%%%%%%%%%%%%%%%%%%%%%%%%%%%%%%%%%%%%%
\begin{figure}[t!]
\includegraphics[width=\columnwidth]{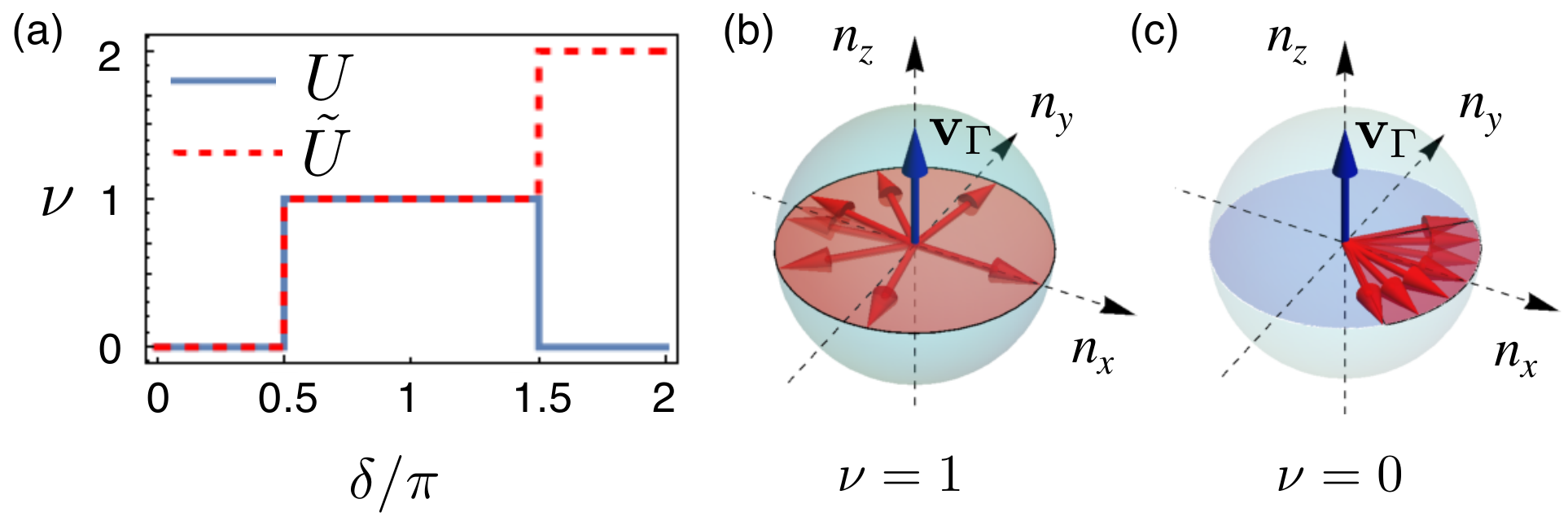}
\caption{{\bf Quantum walk winding numbers} (a) Dependence of the winding number on the parameter $\delta$ for protocols $U$ and $\tilde U$, as indicated in the legend. (b)-(c) On the Poincar\`e sphere we plot eigenstates $\bn(q)$ of the evolution operator $U$, in two cases $\delta=\pi$
and $\delta=\pi/4$, which have winding 1 and 0, respectively. These states are positioned on a single plane, perpendicular to the vector $\mathbf{v}_\Gamma$  ($\Gamma=\mathbf{v}_\Gamma\cdot\bsigma$).}
\label{fig:winding}
\end{figure}
%%%%%%%

A quantum walk is defined in terms of the operator $U$ describing the evolution of a single time-step.
In our experiments we consider the two protocols $U(\delta)=T(\delta)W$ and $\tilde U(\delta)=\sqrt {T(\delta)} W \sqrt {T(\delta)}$.
Both feature chiral symmetry, with their winding number $\nu$ depending on $\delta$ as shown in the phase diagrams in Fig.\ \ref{fig:winding}(a)~\cite{Cardano2017}.
In Fig.\ \ref{fig:winding}(b),(c) we plot the eigenstates $\bn(q)$ of protocol $U$, in two illustrative cases with $\delta=\pi$ and $\delta=\pi/4$.
Vectors $\pm\mathbf{v}_{\Gamma}$ individuate the eigenstates of the chiral
symmetry operator $\Gamma$, that we refer to as $\ket +$ and $\ket -$, respectively. 

%%%%%%%%%%%%%%%%%%%%%%%%%%%%%%%%%%%%%%%
% Figure: setup
%%%%%%%%%%%%%%%%%%%%%%%%%%%%%%%%%%%%%%%
\begin{figure}[t!]
\includegraphics[width=\columnwidth]{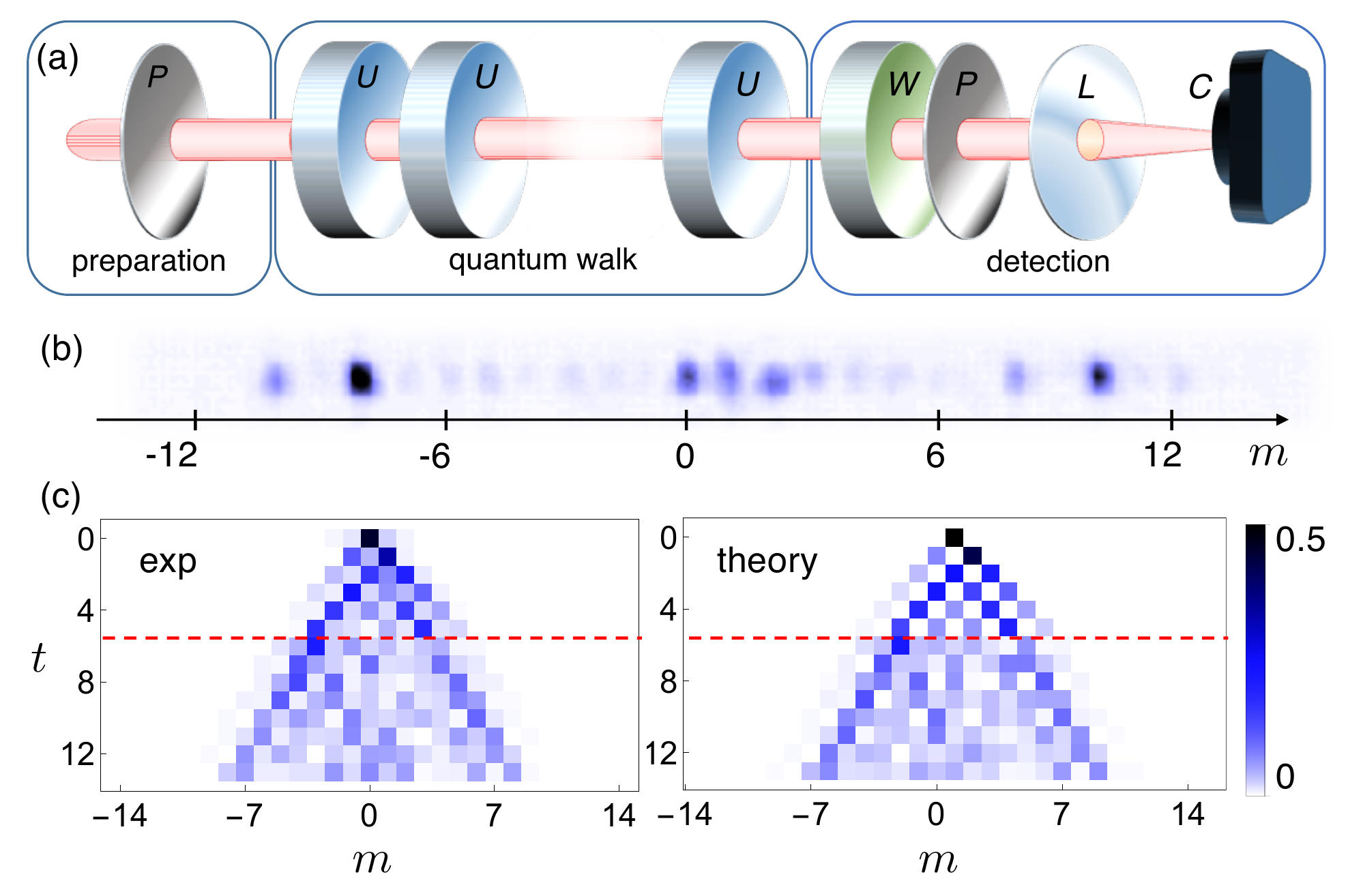}
\caption{{\bf Experimental setup and chiral probability distributions.} 
(a) In our experiment the input beam is prepared with a polarizer ($P$), then it performs a QW generated by either $U$ or $\tilde U$ (see main text). 
Finally, the chiral polarization components are analyzed with a QWP (W) and a polarizer, and the light intensity is recorded on a camera (C) placed in the focal plane of a lens. 
(b) Representative intensity distribution recorded by the camera. (c) Probability distributions $P_-(m)$ for a QW with a quench (at the 5th step, see the red dashed line) from the chiral protocol $U(\pi)$ to the chiral protocol $U(2\pi/5)$. Experimental results (left) are compared to numerical simulations (right). 
}
\label{fig:setup}
\end{figure}
%%%%%%%%%%%%%%%%%%%%%%%%%%%%%%%%%%%%%%%

\begin{figure}[t!]
\centering
\includegraphics[width=\columnwidth]{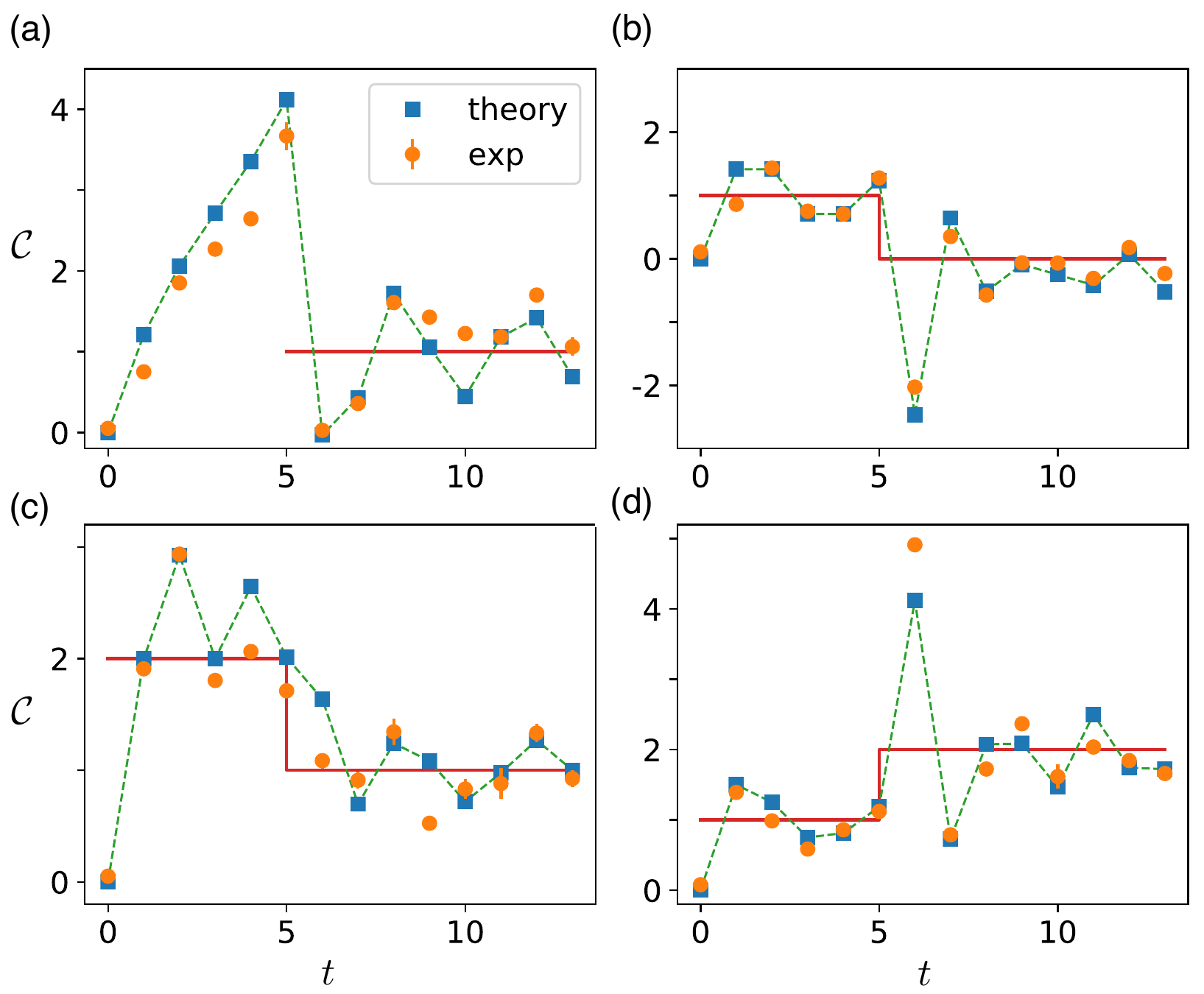}
\caption{\label{fig:experiments}
{\bf MCD in quenched QWs.} MCD in quantum walks governed by protocol $V_1$ from step 1 to 5, and by protocol $V_2$ from 6 to 13. 
The winding number $\nu$ of each protocol is shown with red lines.
(a) $V_1=T(3\pi/4)S$ is non-chiral (therefore its winding is not defined), $V_2=U(\pi)$.
(b) $V_1=U(\pi)$, $V_2=U(2\pi/5)$. 
(c) To study quenches between non-trivial phases, we chose here $V_1=\tilde U(7\pi/4)$ and $V_2=\tilde U(\pi)$, having winding numbers 2 and 1, respectively. (d) Initial and final evolutions are swapped with respect to panel (c). 
In all plots, experimental data (orange dots) are compared with theoretical simulations (blue squares). In panels (c)-(d), an extra plate implementing the operator $T[(\delta_2 -\delta_1)/2]$ was inserted between the $5^{\rm th}$ and $6^{\rm th}$ steps.
}
\end{figure}

These QWs are simulated in the setup sketched in Fig.\ \ref{fig:setup}(a). A laser-light beam (wavelength $\lambda=632$ nm) with  beam waist $\omega_0\simeq 5$ mm propagates through a sequence of QWPs and $g$-plates, arranged so as to realize either protocol $U$ or $\tilde U$. Before entering the QW setup, the optical field corresponds to the spatial mode $\ket{m=0}$, thus realizing the localized initial conditions given in the definition of the MCD (see Eq.\ \eqref{eq:generalMCD}). A polarizer guarantees that its polarization is horizontal. 
At the exit of the walk a QWP and a polarizer, both mounted on rotating mounts, allow us to analyze individual polarization components associated with
$\ket +$ and $\ket -$. After the polarization analysis stage, we place a lens and a we place a lens and a camera to record light intensity in the focal plane. 
Here, the field intensity distribution features spatially separated Gaussian spots, each corresponding to a mode $\ket m$ (see Fig.\ref{fig:setup}(b)). The overlap between the modes in the lens focal plane can be adjusted by tuning the spatial period of the gratings $\Lambda$. 
Choosing $\Lambda\simeq\omega_0=5$ mm is sufficient to make this overlap negligible \cite{DErrico2020}. Recorded intensity patterns are processed to extract the particle probability distributions $P_\pm(m)$ associated with chiral polarization states $\ket\pm$, respectively (see Fig.\ \ref{fig:setup}(c)). We measure these distributions at each timestep $t$, we compute the mean value of their difference, and we obtain the MCD defined in Eq.~\eqref{eq:generalMCD}:
%%%%%%
\begin{align}\label{eq:MCD}
\C(t)=\sum_m 2\,m\left[P_+(m,t)-P_-(m,t)\right].
\end{align}
As discussed above, in our QW (having a unit cell with $\D=2$) the sum in Eq.~\eqref{eq:generalMCD} can be replaced by a single measurement, considering an arbitrary input state localized on a single unit cell at $t=0$. In our experiments, the polarization state of the light beam at the QW entrance is $(\ket{L}+\ket{R})/\sqrt 2$, that is a horizontal polarization. The first experimental validation of Eq.\ \eqref{eq:main} is obtained by using the first five steps of our platform to implement the nonchiral protocol $T(3 \pi/4)\cdot S$, where $S=\sigma_x$ is the operator associated with a half wave plate. The remaining part of the setup is arranged so as to perform eight steps of the QW $U(\pi)$. The results are shown in Fig.\ \ref{fig:experiments}(a). We observe that after few timesteps the MCD oscillates around $\nu$, with an amplitude getting smaller as $t$ increases.

%%%%%%%%%%%%%%%%%%%%%%%%%%%%%%%%%%%%%%%
% Figure: experiments on MCD
%%%%%%%%%%%%%%%%%%%%%%%%%%%%%%%%%%%%%%%
%%%%%%%%%%%%%%%%%%%%%%%%%%%%%%%%%%%%%%%

Finally, we perform experiments to study quenches between different topological phases of both $U$ and
$\tilde U$ protocols. In  Fig.\ \ref{fig:experiments}(b)-(d) we show the evolution of the MCD in QWs where at the 6th step the value of $\delta$
has changed. In doing so, the system evolves under evolution operators associated with different winding numbers (see the figure caption for more details). All the experimental data correctly reproduce the oscillating behavior featured by theoretical simulations, showing that in a quench architecture the detection of 
the MCD allows one to monitor the winding number, faithfully signaling dynamical phase transitions. Error bars are the m.s.e.\ obtained from a set of 4 repeated measurements, each experiment being performed by re-aligning all plates in the QW setup. A few experimental points lie more than three standard deviations away from values simulated numerically. This is ascribed to the presence of imperfections, such as defects in our plates, that induce deviations which are systematic and hence not taken into account in our experimental estimate of statistical uncertainties. The number of QW steps performed in this experiment was mainly limited by the number of liquid-crystal plates at our disposal, yet it was sufficient to provide a clean experimental demonstration of our main finding, contained in Eq.\ \eqref{eq:main}. In principle, one could reach a larger number of steps, provided that the condition $m\Delta k_\perp\ll k_z$ holds true for each mode $\ket{m}$. In our case, practical issues were the limited transmittance of the liquid-crystal plates, which can be improved with dedicated anti-reflection coatings, and the additional relative phase shifts accumulated by different modes during free-space propagation. These could be eliminated by introducing suitable imaging systems between consecutive steps. Further details can be found in Ref. \cite{DErrico2020}.

\section{Conclusions and outlooks}
We have shown that the mean chiral displacement is a powerful tool to probe the topology of chiral 1D systems whose initial state is connected to a localized one via a unitary and translation-invariant transformation. As such, the MCD can be used as a topological marker in experiments where the underlying Hamiltonian is suddenly quenched between different topological phases, like those studying topological systems out of equilibrium and dynamical topological phase transitions \cite{Wang2018c,Heyl2019,Tarnowski2019}. 
Remarkably, we showed that the MCD always performs damped oscillations around the winding number of the instantaneous Hamiltonian. 
In the future, it would be interesting to apply similar ideas to systems with higher dimensional internal states and to non-unitary processes \cite{Yokomizo2019,Longhi2019,Gong2018}.

\section*{Acknowledgments}
AD'E, RB, LM and FC acknowledge financial support from the European Union Horizon 2020 program, under European Research Council (ERC) grant no. 694683 (PHOSPhOR). ADa and ML acknowledge financial support from the Spanish Ministry MINECO (National Plan 15 Grant: FISICATEAMO No. FIS2016-79508-P, SEVERO OCHOA No. SEV-2015-0522, FPI), European Social Fund, Fundaci\'o Cellex, Generalitat de Catalunya (AGAUR Grant No. 2017 SGR 1341 and CERCA/Program), ERC AdG NOQIA, and the National Science Centre, Poland-Symfonia Grant No. 2016/20/W/ST4/00314. ADa is financed by a Juan de la Cierva fellowship (IJCI-2017-33180). PM acknowledges support by the ``Ram\'on y Cajal" program, and by the Spanish MINECO (FIS2017-84114-C2-1-P). ADa, ML and  PM acknowledge support from EU FEDER Quantumcat.

\end{document}